\documentclass[12pt]{article}

\usepackage{scicite}
\usepackage{amsmath,amsthm}
\usepackage{graphicx}
\usepackage{graphics}
\usepackage{float}
\usepackage{subfig}
\usepackage{caption}
\usepackage{hyperref}
\usepackage{times}

\newenvironment{sciabstract}{%
\begin{quote} \bf}
{\end{quote}}

\title{Satellite Laser-Ranging as a Probe of Fundamental Physics}

\author
{Ignazio Ciufolini,$^1$ Richard Matzner,$^{2\ast}$  Antonio Paolozzi,$^3$\\ 
Erricos C. Pavlis,$^4$ Giampiero Sindoni,$^3$ John Ries,$^5$\\
 Vahe Gurzadyan,$^6$  Rolf Koenig,$^7$\\
\\
\normalsize{$^{1}$Dip. Ingegneria dell'Innovazione, Universit\`a del Salento, }\\
\normalsize{ Lecce, and Centro Fermi, Rome, Italy}\\
\normalsize{$^{2}$Theory Center, University of Texas at Austin, USA}\\
\normalsize{$^{3}$Scuola di Ingegneria Aerospaziale, Sapienza Universit\`a di Roma, Italy}\\
\normalsize{$^{4}$Goddard Earth Science and Technology Center (GEST), }\\
\normalsize{University of Maryland, Baltimore County, USA}\\
\normalsize{$^{5}$Center for Space Research, University of Texas at Austin, USA}\\
\normalsize{$^{6}$Center for Cosmology and Astrophysics, Alikhanian National Laboratory and }\\
\normalsize{ Yerevan State University, Yerevan, Armenia}\\
\normalsize{$^{7}$Helmholtz Centre Potsdam German Research Centre for Geosciences - GFZ, Germany}\\
\\
\normalsize{$^\ast$To whom correspondence should be addressed; E-mail: richard.matzner@sbcglobal.net.}
}

\date{}

\begin{document} 


\baselineskip24pt


\maketitle


\begin{sciabstract}
Satellite laser-ranging is successfully used in space geodesy, geodynamics and Earth sciences; and to test fundamental physics and specific features of General Relativity. We present a confirmation to approximately one part in a billion of the fundamental weak equivalence principle (``uniqueness of free fall'') in the Earth's gravitational field, obtained with three laser-ranged satellites, at previously untested range and with previously untested materials. The weak equivalence principle is at the foundation of General Relativity and of most gravitational theories.
\end{sciabstract}


\section*{Introduction}

General Relativity (GR) describes gravitational interaction via the geometry of spacetime whose dynamical curvature is determined by the distribution and motion of mass-energy; concurrently the motion of mass-energy is determined by the spacetime geometry. ``Mass tells spacetime how to curve and spacetime tells mass how to move'' (Wheeler, \cite{bib1}). However, for such a geometrical picture to work, any two particles, independently of their mass, composition and structure, must follow the same geometrical path of spacetime \cite{bib2,bib6,bib7}. The weak equivalence principle states that the motion of any test particle due to the gravitational interaction with other bodies is independent of the mass, composition and structure of the particle. [A test particle is an electrically neutral particle, with negligible gravitational binding energy, negligible angular momentum and small enough that the inhomogeneities of the gravitational field within its volume have negligible effect on its motion.] Thus, the motion of planets, stars, and galaxies in the universe is simply dictated by the geometry of spacetime: they all follow purely geometrical curves of the spacetime called geodesic \cite{bib1,bib2,bib5,bib4}. A geodesic is the generalization to a curved spacetime of a straight line of the flat Euclidean geometry. 
[The surface of a sphere is an example of a non-Euclidean geometry with positive curvature.]
For example the motion of the Moon due to the Earth is not determined by the gravitational force that the Earth's mass exerts on the Moon as in Newtonian theory. Rather the Moon
is simply following a geometrical curve in spacetime, a geodesic, independent of its properties such as mass, composition, and structure, depending only on its initial conditions of position and velocity \cite{bib5}. Then, for example, the observed (approximately) elliptical orbit of the Moon around the Earth is just the projection to our three dimensional space of the geodesic followed by the Moon in the four-dimensional curved spacetime geometry generated by the Earth's mass (see Fig. 1a). 

There are a number of different formulations of the equivalence principle. The weak equivalence principle, also known as the Galilei equivalence
 principle, is based on the principle that the ratio of the inertial mass to the passive gravitational mass is the same for all bodies. This last formulation is also known as the Newton equivalence
 principle. The weak form is at the basis of most known viable theories of gravity. The medium form states that locally, in freely falling frames, all the non-gravitational laws of physics are the laws
 of special relativity \cite{bib4}; the strong form includes gravitation itself in the local laws of physics, meaning that an external gravitational field cannot be detected in a freely falling frame by its
 influence on local gravitational phenomena. The medium form is at the basis of any gravitational theory based on a spacetime geometry described by a symmetric metric tensor, the so-called
 metric theories of gravitation, and the strong form is a cornerstone of GR. Since the weak equivalence principle underlies the geometrical structure of GR as well as our understanding of the dynamics of the universe and of
 astrophysical bodies, it has been tested in very accurate experiments \cite{bib2,bib6,bib7}. Its tests go from the pendulum experiments (and inclined tables) of Galileo Galilei (about 1610),
 Christian Huygens (1673), Isaac Newton (1687) and Bessel (1832), to the classic torsion balance experiments of E\"{o}tvos \cite{bib8} (1889 and 1922) in the gravitational field of Earth (at a range
 from the center of $\sim$ 6370 km). Roll, Krotkov and Dicke \cite{bib9} (1964) used aluminum and gold in the gravitational field of Sun (at a range of $\sim 1.5 \cdot$ 10$^{8}$ km) with a
 precision of $\sim 10^{-11}$, and Braginsky and Panov \cite{bib10} (1972) used aluminum and platinum in the gravitational field of Sun with a precision $\sim 10^{-12}$. The 2012 test by the
 University of Washington \cite{bib11} (the so-called ``Eot-Wash" experiment)  used a torsion balance to confirm the weak equivalence principle for beryllium-aluminum and beryllium-titanium test
 bodies in the field of the Earth to a precision of $\sim 10^{-13}$. In April 2016 the space experiment {\it MICROSCOPE} of CNES (Centre National d'\'{E}tudes Spatial) was successfully inserted
 into orbit at an altitude of approximately 711 km. It was designed to test the equivalence principle to a precision of $\sim 10^{-15}$ comparing the motion of two proof masses, one of titanium-aluminum alloy, and one of a platinum-rhodium alloy. The science phase of the mission lasted for about two years.  First results (agreement with the weak equivalence principle to parts in $10^{14}$) were published in December 2017; the final
measurements are expected in Summer, 2019 \cite{bib12}. Additional novel tests include \cite{bib12a}, based on the time coincidence of gravitational radiation with electromagnetic observations
 from the LIGO event GW170817. Also, \cite{bib12c}, \cite{bib12d}  discuss the equivalence principle in the context of quantum systems.

Remarkably, GR even incorporates the strong equivalence principle \cite{bib1}: gravitational energy (e.g. gravitational binding energy or the effective energy content of gravitational radiation) acts
 as a source (an active gravitational mass) for the gravitational field just like any other mass-energy, and responds to an external gravitational field (falls in that field) like any other passive
 gravitational mass. The strong equivalence principle has been validated by comparing acceleration of the Moon and the Earth toward the Sun using Lunar Laser Ranging, which measures the
 motion of the Moon relative to the Earth at the centimeter level. (The Earth's fractional gravitational binding energy is about twenty times that of the Moon.) Differential acceleration of the two bodies
 would lead to polarization of the Moon's orbit; this has been excluded to parts in 10$^{13}$ by Williams, Turyshev and Boggs \cite{bib13}. Archibald et al. \cite{Archetal} studied the system PSR
 J0337+1715, and a gave a limit on the strong equivalence principle. The system consists of a triple: a tight (1.6 day) millisecond pulsar - white dwarf binary, in 327-day orbit about a distant white
 dwarf. Study shows the accelerations of the pulsar and its nearby white dwarf companion differ fractionally by no more than $2.6 \times 10^{-6}$ as they fall toward the distant white dwarf.

Weak equivalence may be violated if there is a weak (of roughly the same strength as gravity) fundamental field that couples to matter differently from the universality of gravity. For instance, some theoretical constructs suggest an almost massless scalar field which couples to the nucleon number, rather than to the total mass-energy of the object. This scalar gravity would therefore be composition-dependent (thus violating the weak equivalence principle) since the fractional nuclear binding energy is different among elements. Different gravitational theories can exhibit a breakdown of the weak equivalence principle depending on the range, for example compared to the range of Yukawa-type  deviations from the inverse square law of gravitation \cite{bib15} in a theory that couples to nucleon number. A composition dependent interaction between two bodies might be described by the following potential energy of a body 1 in the gravitational field of a body 2:
\begin{equation}\label{eq1}
U(r)=-\frac{G M_{1}M_{2}}{r}\bigg( 1 +\frac{b_{1}b_{2}}{GM_{1}M_{2}}e^{-\frac{r}{\lambda}}\bigg)
\end{equation}
where $- G M_{1} M_{2} /r$ is the standard Newtonian potential energy (representing the Newtonian gravitational theory as the lowest order approximation of GR), $G$ is the gravitational constant, $M_{1}$ and $M_{2}$ are the masses of the two bodies, $b_{1}$ and $b_{2}$ are some composition dependent properties of bodies 1 and 2 defining the additional interaction, $r$ is the distance between the two bodies and $\lambda$  the Yukawa range of the interaction. 

The ratio $b/M$ will in general be different for each body and thus bodies with different compositions will fall with different acceleration, violating the uniqueness of free fall. Furthermore, a measurable deviation from the universality of free fall may depend not only on the material of the proof masses, nucleon number, etc., but also, as in Eq(1), on the range of the experiment (an effective change of $GM_\odot$ with distance) unless $\lambda \approx \infty$. Therefore it is important to test the equivalence principle with different materials and at different ranges; an important aspect of the present determination is the distance scale involved. In our analysis we assume $GM_\odot$ is a universal constant, and cast the problem entirely in terms of the universality (or not) of the ratio $(m_g/m_i)$.

\section*{Test of Equivalence Principle: Laser-Ranged Satellites LARES, LAGEOS, and LAGEOS 2}

We describe a test of the weak equivalence principle using for the first time freely falling high altitude laser-ranged satellites: LARES, made of sintered tungsten \cite{bib22,bib23}; and LAGEOS \cite{bib24} and LAGEOS 2 \cite{LAG}, two almost identical satellites each composed of 57\% aluminum shell / 43\% brass core by mass. These are materials never previously tested.  Further details about the satellites are found in the Section {\bf Methods} below. The number of well tracked dense laser ranged satellites is not large, so if new satellites meeting these criteria are launched their inclusion would improve our analysis to (at least partially) disentangle the weak equivalence result from a more controversial change of $GM_{\bigoplus}$ with distance. [Such a gradient could be the result of a violation of the crucial theorem that a gravitating sphere acts as a point mass (shell theorem). The most general form for the force to fulfill the shell theorem, $F(r)= A r^{-2} + \Lambda r$ \cite{bib21bis} contains a cosmological constant $\Lambda$, which LARES and LAGEOS data constrain.  Constraints on modified gravity laws, including\ Yukawa type, are essential for GR's Newtonian law as limit, as well as for understanding the dynamical features in the local group of galaxies and its vicinity (see, e.g. \cite{bib21ter}).]
                                                                                                                                                                                                                         
The self-gravities  of all three satellites are negligible. By comparing the residual radial accelerations of these three satellites, we obtain a test validating the weak equivalence principle with an accuracy of $\sim 10^{-9}$. The range of the test described here goes from $\sim$7820 km from Earth's center (altitude 1450 km) for the LARES satellite to $\sim$12200 km from Earth's center for LAGEOS and LAGEOS 2. Our test thus fills a distance gap not covered by the laboratory and Lunar Laser Ranging tests; any scale range in principle will constrain parameters entering the ``fifth'' force, phenomenology or coupled gravity models.

\section*{Orbital Analysis and Results}

We processed more than half a million normal points of the three satellites LARES, LAGEOS, and LAGEOS 2. The laser ranging normal points were processed using NASA's orbital analysis and data reduction software GEODYN II \cite{bib25}, and validated by the orbital modelers UTOPIA\cite{utopia}, and EPOSOC\cite{eposoc}. The data analysis was based on the Earth gravity model GGM05S \cite{Ries}. [The Earth gravity field model GGM05S was released in 2013, based on approximately 10 years of data of the GRACE (Gravity Recovery and Climate Experiment) \cite{bib27,bib28}  spacecraft. It describes the Earth's spherical harmonics up to degree 180. The NASA-DLR (Deutsche Zentrum f\"{u}r Luft- und Raumfahrt: the German Aerospace Center) GRACE space mission consists of twin spacecraft launched in a polar orbit at an altitude of approximately 400 km and $\sim 200-250$ km apart. The spacecraft range to each other using radar and are tracked by the global positioning satellites. GRACE has greatly improved our knowledge of the Earth's gravitational field.]
The GEODYN analysis includes Earth rotation from Global Navigation Satellite Systems (GNSS) and Very Long Baseline Interferometry (VLBI), Earth tides, solar radiation pressure, Earth albedo, thermal thrust, and lunar, solar and planetary perturbations. We analyzed the laser ranging data of the LARES, LAGEOS, and LAGEOS 2 satellites from February 2012 to December 2014.  The laser ranging data for LARES, LAGEOS, and LAGEOS 2 were collected from more than 40 ILRS stations all over the world \cite{bib29}.

If we include the acceleration due to the Earth's quadrupole moment (the Earth's oblateness measured by the $J_{2}$ coefficent \cite{bib30}), and the potential breakdown of the uniqueness of free fall, the radial acceleration $a_r$ of an Earth satellite  can be written:

\begin{equation}\label{eq3}
a_r=-\frac{m_{g}}{m_{i}} \frac{GM_{\bigoplus}}{r^{2}}\biggl[1-3J_{2}\biggl(\frac{R_{\bigoplus}}{r}\biggl)^{2}P_{20}+\ldots\biggr].
\end{equation}
[Note that $a_r$ is $not$ the second time derivative $\ddot r$ of the radial coordinate $r$.
Consider for instance circular motion where the radius $r$ is constant, but $a_r = - r \dot\theta^2 $.]
Here, for simplicity, we have included within $m_{g}/m_{i}$  any breakdown of the uniqueness of free fall, for example of the type of the second term of Eq.(1). ($m_{g}$ is the passive gravitational mass of the satellite and $m_{i}$ its inertial mass; $m_{g}/m_{i}$ is is a universal constant in GR and Newtonian Physics, equal to unity by choice of units.) $M_{\bigoplus}$ and $R_{\bigoplus}$ are the Earth's mass and equatorial radius, $r$ is the radial distance of the satellite from the Earth barycenter, and $P_{20}$ is the associated Legendre  function, of degree 2 and order 0, of the satellite latitude (see Methods). The product $GM_{\bigoplus}$ for the Earth is today measured \cite{bib31} to be 398600.4415 km$^{3}$/sec$^{2}$ (including the mass of the atmosphere) with an estimated {\it relative} 
 ($one$-$sigma$)
 uncertainty of $\sim 2 \cdot 10^{-9}$. The Earth's quadrupole moment \cite{bib26} $J_{2}$ is equal to 0.0010826358 with a {\it relative} uncertainty of $\sim 10^{-6}$ to $10^{-7}$.
According to the uniqueness of free fall, the ratio $m_{g}/m_{i}$ is the same for every test body. Here we consider the possibility that such a ratio may be different for aluminum/brass of the LAGEOS satellites, at a distance of $\sim 12220$ km from the Earth center, and for the tungsten alloy of the LARES satellite, at a distance of $\sim 7820$ km from the Earth's center. On the basis of the LAGEOS, LAGEOS 2 and LARES laser-ranging observations, we then set an experimental limit on the deviation $\delta(m_g/m_i)$:

\begin{equation}\label{eq4}
\delta\bigg(\frac{m_{g}}{m_{i}}\bigg)= \frac{m_{g}}{m_{i}}\bigg|_{tungsten} - \frac{m_{g}}{m_{i}}\bigg|_{aluminum/brass}.
\end{equation}
$\delta(GM_{\bigoplus})$, $\delta J_{2}$, $\delta(m_{g}/m_{i})$ and the measurement error $\delta r$ of the radial distance, $r$, of the three satellites are the main uncertainties in  our estimation of the radial accelerations, Eq.(3), of the three satellites. See Eq.(9) below.

The radial accelerations of the three satellites are  modeled with our orbital estimator GEODYN II \cite{bib25} using the nominal (fiducial) values of $GM_{\bigoplus}$, $J_{2}$ and $m_{g}/m_{i}$ = 1 and the measured Earth-station to satellite distances. The observed-minus-modeled radial accelerations are computed for every five-day period. These residual radial accelerations of LAGEOS, LAGEOS 2 and LARES are shown respectively in Figs. 2a, 2b, and 2c. The variations around the mean in these figures are due to the uncertainties in the deviations of the Earth's gravity field from spherical symmetry, i.e. to the uncertainties in higher Earth spherical harmonics. The residuals of LAGEOS shown in Fig. 2a are smaller than those of the other two satellites since the value of $GM_{\bigoplus}$ used in GEODYN was obtained \cite{bib31} using the LAGEOS laser-ranging data. We normalize $m_g/m_i =1$ for LAGEOS and the essentially identical (in both composition and altitude) LAGEOS 2. Thus $\delta (m_g/m_i)$ can only appear in consideration of LARES. Eqs. (9) and (10) below give the relations between the variations of the accelerations, and the parameter variations.
The long term average residual accelerations for LARES are comparable to those for LAGEOS 2 even though LARES orbits at a much lower altitude (Fig. 2c) and LARES undergoes larger single-point excursions.

We observe the average residual radial accelerations:
\begin{eqnarray}
<\delta {a_r}>_{LAGEOS} &=& - 4.056  \cdot10^{-10} m/s^{2} \\
 \nonumber
<\delta {a_r}>_{LAGEOS2} &=& - 2.217  \cdot10^{-9} m/s^{2} \\
 \nonumber
<\delta {a_r}>_{LARES}  &=& + 2.834  \cdot10^{-9} m/s^{2}
\end{eqnarray}
%
%
%
%
%
Here the angle brackets ``$<$ $>$" are long term averages.

The satellites' residual radial accelerations are mainly due to the errors $\delta(GM_{\bigoplus})$, $\delta J_2$, $\delta(m_{g}/m_{i}) $ {\it and the measurement error in the radial distance, $\delta r$, of the three satellites}. Recent studies of the best station performance in ranging to the LAGEOS and LARES satellites suggest $one$-$sigma$ $\delta r \sim \, 2 \, mm$ for the LAGEOS satellites and $\delta r \sim \, 3 \, mm$ for LARES, which we adopt. 

The method is to take the three equations for the residual radial acceleration of each of the three satellites (e.g., in Eq.(11) in {\bf Methods} for LARES). These can be viewed as giving a vector of radial accelerations (Eq.(4)) equal to a square matrix {\bf M} times a vector of
measurement uncertainties: ($ \delta(m_{g}/m_{i}) $, $\delta(GM_{\bigoplus})$, $\delta J_2$), plus (a term proportional to radial measurement uncertainty , $+$ {\it other errors}). We invert this equation (multiply by {\bf M}$^{-1}$), which yields ({\bf Methods})):
$$
\begin{pmatrix}
\delta(m_g/m_i)_{LARES}\\
\delta(GM_{\bigoplus})/GM_{\bigoplus}\\
\delta J
\end{pmatrix}
=
\begin{pmatrix}
2.0 \times 10^{-10}   \\
7.3 \times 10^{-10}  \\
4.3 \times 10^{-9}
\end{pmatrix}
+
\begin{pmatrix}
\pm 1.1  \times 10^{-9}   \\
\pm 2.9  \times 10^{-10}  \\
\pm 3.0  \times 10^{-9}
\end{pmatrix}
$$

The column vector on the left represents the ``decoupled" deviations of $(m_g/m_i)_{LARES}$, $GM_{\bigoplus}$, and $J_2$ from their nominal values. In particular $\delta(m_{g}/m_{i}) $ is independent of the uncertainties $\delta(GM_{\bigoplus})$ and $\delta(J_{2})$. The last column vector on the right represents the uncorrelated $\delta r$ measurement errors; they turn out to dominate our result for $\delta (m_g/m_i)$. [Ranging to laser ranged satellites involves errors arising from atmospheric effects, photon statistics, and geometrical errors because the return comes from a retroreflector array which is not at the center of mass of the satellite.] Thus $\delta (m_{g}/m_{i})$ is determined up to average residuals arising from random $\delta r$ errors from the three satellites, and other smaller errors. See {\bf Methods}.
The resulting value of the deviation $\delta(m_{g}/m_{i})$  for \textit{tungsten} and \textit{aluminum/brass}, Eq.(3), is:
\begin{equation}\label{eq8}
\delta (m_{g}/m_{i})= 2.0 \times 10^{-10} \pm 1.1  \times 10^{-9}
\end{equation}
where $\pm 1.1 \cdot10^{-9}$ is the estimated systematic error principally due to the error in the measurement of the radial distance. Uncertainties in the modeling of the radial accelerations due to the errors in the Earth's spherical harmonics higher than the quadrupole moment, $J_{2}$, and due to the errors in the modeling of atmospheric drag and of other non-gravitational perturbations, such as direct solar radiation pressure and Earth albedo, are included in the {\it other errors} above, and are much smaller. The combined residuals affecting $\delta(m_{g}/m_{i})$ are shown in Fig. 3. Eq.(5) shows a confirmation of the equivalence principle for the three satellites with an accuracy of $\sim \pm 10^{-9}$.

\section*{ Discussion}

Our test of the weak equivalence principle (uniqueness of free fall) using the laser ranged satellites LARES, LAGEOS and LAGEOS 2 fills a gap in the tests of this principle fundamental to Einstein's gravitational theory of General Relativity. 

Some alternative theories of gravitation predict deviations from the uniqueness of free fall that are enhanced at certain ranges depending on a typical scale length and are enhanced for different materials. 
Previous tests of the weak equivalence principle were Earth laboratory tests in the gravitational field of Earth (at a distance of $\sim 6370$ km from the center of the Earth) and at the MICROSCOPE distance of $\sim 7000$ km, and Earth laboratory tests in the gravitational field of the {\it Sun} (at  $\sim 1.5 \times$ 10$^{8}$ km), there were no tests at a range between 7820 km and 12270 km prior to the present test using the LAGEOS and LARES satellites. Furthermore the uniqueness of free fall was never previously confirmed comparing test bodies made of aluminum/brass and tungsten, such as the LAGEOS and LARES satellites. Our test has confirmed the validity of the weak equivalence principle for these metals over ranges 7820 km and 12270 km, to accuracy of one part per billion.
Also, since LARES differs both in composition and in orbital radius from LAGEOS and LAGEOS 2 (which are very similar to one another in these properties) our observation can be viewed as constraining $(\delta G)/G$,  again to $\sim$ one part in $10^{-9}$, over the range $7820$ to $12270 \, km$.

\section*{Methods}

The orbits \cite{bib22} of the three satellites are almost circular, with very small orbital eccentricity. The LAGEOS satellite has semimajor axis 12270 km (altitude 5890 km), orbital eccentricity of 0.0045 and orbital inclination of $109.84^{\circ}$; LAGEOS 2 has semimajor axis 12160 km (altitude 5780 km), orbital eccentricity of 0.0135 and orbital inclination of $52.64^{\circ}$; LARES, semimajor axis 7820 km (altitude 1530 km), orbital eccentricity of 0.0008 and orbital inclination of $69.5^{\circ}$. In this analysis we assume all three satellites are in circular orbits. All three are passive, spherical, laser-ranged satellites. LAGEOS was launched in 1976 by NASA, and LAGEOS 2 in 1992 by ASI, the Italian Space Agency, and NASA. They are two almost identical spherical passive satellites covered with corner cube reflectors to reflect back laser pulses emitted by the stations of the satellite laser ranging (SLR) network of the International Laser Ranging Service (ILRS) \cite{bib29}. SLR allows measurement of the radial position of the LAGEOS satellites with a median accuracy of the order two millimeters over an Earth-surface to satellite distance of $\sim 6000$ km. LARES is a satellite of ASI, launched in 2012 by ESA, the European Space Agency, with the new launch vehicle VEGA.  LARES was designed to approach as closely as possible an ideal test particle \cite{bib23}. This goal was achieved by minimizing the surface-to-mass ratio of the spherical satellite (the smallest of any artificial satellite), by reducing the number of its parts, by avoiding any protruding components, and by using a non-magnetic material. LARES carries laser retro-reflectors similar to those on LAGEOS and LAGEOS 2.
Since LARES is at lower altitude, it is accessible to more ranging stations. Some of these stations have slightly reduced timing accuracy (compared to those that range to LAGEOS). As a result, the median accuracy of positioning of LARES is at the roughly the three millimeter level.

The classical gravitational potential of a spheroid, such as the Earth \cite{bib30}, can be written:
\begin{equation}\label{eq9}
  U=\frac{GM_{\bigoplus}}{r}\biggl[1-J_{2}\biggl(\frac{R_{\bigoplus}}{r}\biggl)^{2}P_{20}+\ldots\biggr]
\end{equation}
where $M_{\bigoplus}$ is the Earth's mass, $R_{\bigoplus}$ its equatorial radius, $J_{2}$ its quadrupole moment, $G$ the gravitational constant, $r$ is the radial distance from the origin and $P_{20}$ is the Legendre associated function of degree 2 and order 0:
\begin{equation}
P_{20}(sin\phi)=\frac{3}{2}sin^{2}\phi -\frac{1}{2}
\end{equation}
where $\phi$ is the latitude and in the expression (6) we have neglected higher order $P_{n0}$ terms. 

If there is a violation of the weak equivalence principle, the ratio $m_{g}/m_{i}$ of the gravitational mass to the inertial mass may be different between the aluminum/brass of the LAGEOS satellites and the tungsten alloy of the LARES satellite. Furthermore a composition dependent interaction between two bodies may depend on the distance between the two bodies and on the range of the interaction as described by Eq.(1). We have then indicated with $\delta(m_{g}/m_{i})$ in Eq.(2) any breakdown of the uniqueness of free fall including, for example, one of the type of the second term of Eq.(1). The radial acceleration of a satellite, such as LAGEOS and LARES, can thus be written by Eq.(2) above. Therefore, the leading terms of the \textit{residual, instantaneous} unmodeled radial accelerations of the LAGEOS and LARES satellites can be written:

\begin{equation}\label{eq11}
\delta a_r \cong - \frac{GM_{\bigoplus}}{r^{2}}\; \delta\bigg(\frac{m_{g}}{m_{i}}\bigg)  - \frac{\delta(GM_{\bigoplus})}{r^{2}}+ 3 \frac{GM_{\bigoplus}}{r^{2}}\bigg(\frac{R_{\bigoplus}}{r}\bigg)^{2} P_{20}\; \delta J_{2}  + 2 \frac{GM_{\bigoplus}}{r^{3}} \delta r
\end{equation}
(one equation for each satellite).

We now set $({m_{g}/m_{i}}|_{aluminum/brass}) \equiv 1$ at the altitude of the LAGEOS satellites, but allow nonzero $\delta({m_{g}/m_{i}})$ for the sintered tungsten, lower orbiting LARES. Nonzero $\delta({m_{g}/m_{i}})$ indicates a violation of the equivalence principle between the LARES and the other satellites. Thus the $\delta({m_{g}/m_{i}})$  is possibly nonzero in Eq.(8) only for LARES.

In this formulation we assume a universal value of $GM_{\bigoplus}$ (the same value for all satellites). However we allow a possible offset $\delta (GM_{\bigoplus})$ between its observed, fiducial value, and its true value.  (Hence $\delta (GM_{\bigoplus})$ is itself universal.) Similarly, we assume a universal value of $J_{2}$, with a possible offset $\delta J_2$ between its observed, fiducial value, and its true  value; $\delta J_2$ is also universal.

The meaning of $\delta r$ is different. It is the mean value of the uncertainty in the radial distance of each satellite from the Earth center of mass, mainly due to errors in the determination of the Earth center of mass, biases in laser ranges, errors in the modeling of the dispersion of the laser pulses by the troposphere, and uncertainties arising from determining the precise position of the retroreflector with respect to the center of mass of the satellite. Since these are mean values of uncorrelated errors, there are different values of $\delta r$ for each satellite. We take $3 \, mm$ for LARES and $2 \, mm$ for the LAGEOS satellites.

With Eq.(8), we can write for LAGEOS (here angle brackets ``$<$ $>$" are {\it long term} averages):

\begin{multline}\label{eq12}
<\delta a_r>_{LAGEOS} \cong -
\frac{\delta(GM_{\bigoplus})}{r_{LAGEOS}^{2}}+ 3 \frac{GM_{\bigoplus}}{r_{LAGEOS}^{2}}\bigg(\frac{R_{\bigoplus}}{r_{LAGEOS}}\bigg)^{2} \cdot \\ \cdot \bigg(\frac{3}{4} sin^{2}I_{LAGEOS}-\frac{1}{2}\bigg)\delta J_{2} +
\frac{2GM_{\bigoplus}}{r_{LAGEOS}^{3}}\delta r_{LAGEOS}.
\end{multline}
The coefficient of $\delta J_2$ in Eq.(9) is the average over an orbit of the $P_{20}$  Lagrange associated function of $sin \phi$, where $\phi$ is the latitude of the satellite. This function  can be written as a function of the orbital inclination $I$ and the true anomaly $f$:

\begin{displaymath}
sin\phi=sinI\cdot sin f.
\end{displaymath}

Therefore the average value of $P_{20}$ over one orbital period is:

\begin{displaymath}
< P_{20}> = \frac{\int_0^{2\pi} \big(\frac{3}{2}sin^{2}I \cdot  sin^{2}f-\frac{1}{2}\big)df}{2\pi}=\frac{3}{4}{\sin}^{2}{I}-\frac{1}{2},
\end{displaymath}
which is used in Eq.(9).

A similar expression to Eq.(9) holds for LAGEOS 2. However for LARES we take into account a possible deviation of $m_g/m_i$ from unity, so we have an additional term proportional to $\delta (m_g/m_i)$:
\begin{multline}\label{eq13}
<\delta a_r>_{LARES} \cong
-\frac{\delta(\frac{m_{g}}{m_{i}})(GM_{\bigoplus})}{r_{LARES}^{2}}
-\frac{\delta(GM_{\bigoplus})}{r_{LARES}^{2}}
+ 3\frac{GM_{\bigoplus}}{r_{LARES}^{2}}\bigg(\frac{R_{\bigoplus}}{r_{LARES}}\bigg)^{2} \cdot\\
\cdot\bigg(\frac{3}{4} sin^{2}I_{LARES}-\frac{1}{2}\bigg)\,  \delta J_{2} +
\frac{2GM_{\bigoplus}}{r_{LARES}^{3}} \delta r_{LARES}
\end{multline}

Start with Eq.(4) which gives an {\it observed} column matrix $\bigg[<\delta a_r >\bigg]$ of $<\delta a_r>$ values, $(-4.056 \times 10^{-10}, -2.217 \times 10^{-9}, 2.834\times 10^{-9})m/sec^2$ for LAGEOS, LAGEOS 2, LARES, in that order. Work with normalized (fractional) quantities, so define a column matrix  $\frac{<\delta a_r>}{GM_{\bigoplus}/r^2}$, which normalizes each residual acceleration by the nominal acceleration at that radius. Rewrite the long time averages of Eqs. (9) and (10) above in terms of the normalized residual accelerations:

\begin{equation}
\frac{<\delta a_r>}{GM_{\bigoplus}/r^2} = -  \delta\bigg(\frac{m_{g}}{m_{i}}\bigg)  - \frac{\delta(GM_{\bigoplus})}{GM_{\bigoplus}}+ 3 \bigg(\frac{R_{\bigoplus}}{r}\bigg)^{2} <P_{20}>\; \delta J_{2}  + \bigg[2 \frac{\delta r}{r}  + other\bigg].\\
\end{equation}

The $\delta r$ and ALL the other errors from ranging are grouped into the last bracketed term. The {\it other} terms are small compared to $\delta r$ and will be dropped.  With the assumption that the $\frac{\delta r}{r}$  and  $other$ terms are uncorrelated, they should enter isotropically into this normalized equation.\\

Now, ignore for the moment the $\big[\frac{2\delta r}{r} + other\big]$ terms. Then Eq. (11) is of the form \\

\begin{equation}
\bigg[ \frac{<\delta a_r>}{GM_{\bigoplus}/r^2} \bigg] =
\bigg[{\bf M} \bigg] \bigg[ fractional~ changes \bigg]
\end{equation}

\noindent where $\bigg[ \frac{<\delta a_r>}{GM_{\bigoplus}/r^2} \bigg]$ is the column matrix of these quantities for LAGEOS, LAGEOS 2, and LARES; $\bigg[ fractional~ changes \bigg]$ is the column matrix $[\delta(m_g/m_i)$, $\frac{\delta(GM_{\bigoplus})}{GM_{\bigoplus}}$, $\delta J_{20}]$; and $\bigg[ {\bf M} \bigg]$ is the matrix of coefficients.\\

We need to compute the orbit mean of $P_{20}$ for each satellite, which is ${\frac{1}{2}}(\frac{3}{2} sin^2 I -1)$  $= 0.164,$ $-0.026$, $ 0.158,$ for LAGEOS, LAGEOS 2, and LARES respectively.

Also, $(R_{\bigoplus}/r)^2 = 0.27$ for LAGEOS and LAGEOS 2, and 0.66 for LARES, so the  coefficient of $\delta J_2$ in Eq.(11) is $ C_i = 3 (\frac{R_{\bigoplus}}{r})^{2} <P_{20}>= $  $0.1325,$ $-0.0216,$ $0.308$. These values correspond to $C_1$ (LAGEOS), $C_2$ (LAGEOS 2) and $C_3$ (LARES). We will also need $C_1-C_2 = 0.1541$, $C_2-C_3=-0.3296$, $C_3-C_1= 0.1755$.

As noted, we set $\delta(m_g/m_i)$ to zero for LAGEOS and LAGEOS 2. We also make the approximation that their radii are the same. Then the  matrix in Eq.(12) is
$$
\bigg[{\bf M} \bigg] =
\begin{pmatrix}
0 & -1 & C_1\\
0 & -1 & C_2\\
-1 & -1 & C_3
\end{pmatrix}
\quad\\
$$

Its inverse is \\

$$
\bigg[{\bf M} \bigg]^{-1} = {\frac{1}{C_2-C_1}}
\begin{pmatrix}
C_2-C_3 & C_3-C_1 & C_1-C_2\\
-C_2 & C_1 & 0\\
-1 & 1 & 0
\end{pmatrix}
\quad.\\
$$

Thus
$$
\begin{pmatrix}
\delta(m_g/m_i)\\
\delta(GM_{\bigoplus})/GM_{\bigoplus}\\
\delta J
\end{pmatrix}=
\quad\\
$$

$$
{\frac{1}{C_2-C_1}}
\begin{pmatrix}
C_2-C_3 & C_3-C_1 & C_1-C_2\\
-C_2 & C_1 & 0\\
-1 & 1 & 0
\end{pmatrix}
\quad\\
\Bigg[\begin{pmatrix}
[ \frac{<\delta a_r>}{GM_{\bigoplus}/r^2} ]_1\\
[ \frac{<\delta a_r>}{GM_{\bigoplus}/r^2} ]_2\\
[ \frac{<\delta a_r>}{GM_{\bigoplus}/r^2}]_3
\end{pmatrix}
-2
\begin{pmatrix}
 (\delta r/r)_1\\
 (\delta r/r)_2\\
 (\delta r/r)_3.
\end{pmatrix} \Bigg]
$$
Here we have added back the uncorrelated $\delta r$ terms.\\

 Define $GM_{\bigoplus}= 398600km^3/sec^2$ and express $\delta a_r$ in $km/sec^2$. Then we have the dimensionless quantities \\
$$
\begin{pmatrix}
[ \frac{<\delta a_r>}{GM_{\bigoplus}/r^2} ]_1\\
[ \frac{<\delta a_r>}{GM_{\bigoplus}/r^2} ]_2\\
[ \frac{<\delta a_r>}{GM_{\bigoplus}/r^2}]_3
\end{pmatrix}
=
\begin{pmatrix}
(-4.056 \times 10^{-13} )/[398600/(12270)^2] \\
(-2.217 \times 10^{-12}  )/[398600/(12160)^2 ]  \\
(+2.834 \times 10^{-12}  )/[398600/(7820)^2]
\end{pmatrix}
=
\begin{pmatrix}
-1.53 \times 10^{-10}  \\
-8.22 \times 10^{-10}  \\
+4.35 \times 10^{-10}  
\end{pmatrix}
$$

Then inverting Eq.(12) (and setting the $ \delta r$ terms to zero) gives:\\

$$
\begin{pmatrix}
\delta(m_g/m_i)\\
\delta(GM_{\bigoplus})/GM_{\bigoplus}\\
\delta J
\end{pmatrix}
=
\begin{pmatrix}
+2.0 \times 10^{-10}   \\
+7.3 \times 10^{-10}  \\
+4.3 \times 10^{-9}
\end{pmatrix}
$$

We now address the last term in Eq.(11), the column matrix of $(2\delta r/r)_i$, which gives the magnitude of the uncorrelated errors for the satellites. These are the largest remaining uncontrolled errors.

Consider the $2 \delta r/r$-induced errors in the equivalence principle term:

\begin{equation}
error(\delta(m_g/m_i))=-{\frac{1}{C_2-C_1}} \bigg[ (C_2-C_3)(2\delta r/r)_1 ``+" (C_3-C_1)(2\delta r/r)_2 ``+" (C_1-C_2)(2\delta r/r)_3 \bigg]\\
\end{equation}

We use the quotes on the operators ``$+$'' because in fact the error $\delta r$ is uncorrelated between satellites, so we add these errors by quadrature. Also, though this term has an explicit ``-" sign, it is actually stochastic, so contributes ``$\pm$" to the errors.

Thus the full statement of our result is:

$$
\begin{pmatrix}
\delta(m_g/m_i)\\
\delta(GM_{\bigoplus})/GM_{\bigoplus}\\
\delta J
\end{pmatrix}
=
\begin{pmatrix}
+2.0 \times 10^{-10}   \\
+7.3 \times 10^{-10}  \\
+4.3 \times 10^{-9}
\end{pmatrix}
+
\begin{pmatrix}
\pm 1.1 \times 10^{-9}   \\
\pm 2.9 \times 10^{-10}  \\
\pm 3.0 \times 10^{-9}
\end{pmatrix}
$$
The physical result from these calculations is our statement of the equivalence principle:
$\delta(m_{g}/m_{i})= 2.0 \times 10^{-10} \pm 1.1  \times 10^{-9}$ among the three satellites,  consistent with the result $zero$ to within the $\sim 10^{-9} $ fractional accuracy of the determination.

\section*{Data and materials availability}

The laser-ranging data of LARES, LAGEOS and LAGEOS 2 are available at the NASA's archive of space geodesy data CDDIS (Crustal Dynamics Data Information System) \cite{bib33}.

\section*{Acknowledgments}

We gratefully acknowledge the Italian Space Agency for the support of the LARES and LARES 2 space missions
under agreements No. 2017-23-H.O. and No. 2015-021-R.O. We are also grateful to the International Laser Ranging 
Service (ILRS), ESA, AVIO and ELV. ECP acknowledges the support of NASA Grants NNX09AU86G, NNX15AT34A, and NNX14AN50G. 
RM acknowledges NSF Grant PHY-1620610 and JCR the support of NASA Contracts JPL1478584 and NNG17V105C.

\section*{Additional information}
Competing interests: 
The authors declare no competing interests.

\begin{figure}[htbp]
\centering
\subfloat[Two bodies with the same initial conditions follow the same geodesic of spacetime. The projection of the spacetime geodesic onto a spatial plane is, for example, an ellipse (with suitable coordinates). Here, the third spatial dimension is suppressed and the much smaller relativistic precession of the pericenter is not shown.\label{fig:1a}]{\includegraphics[width=0.45\textwidth]{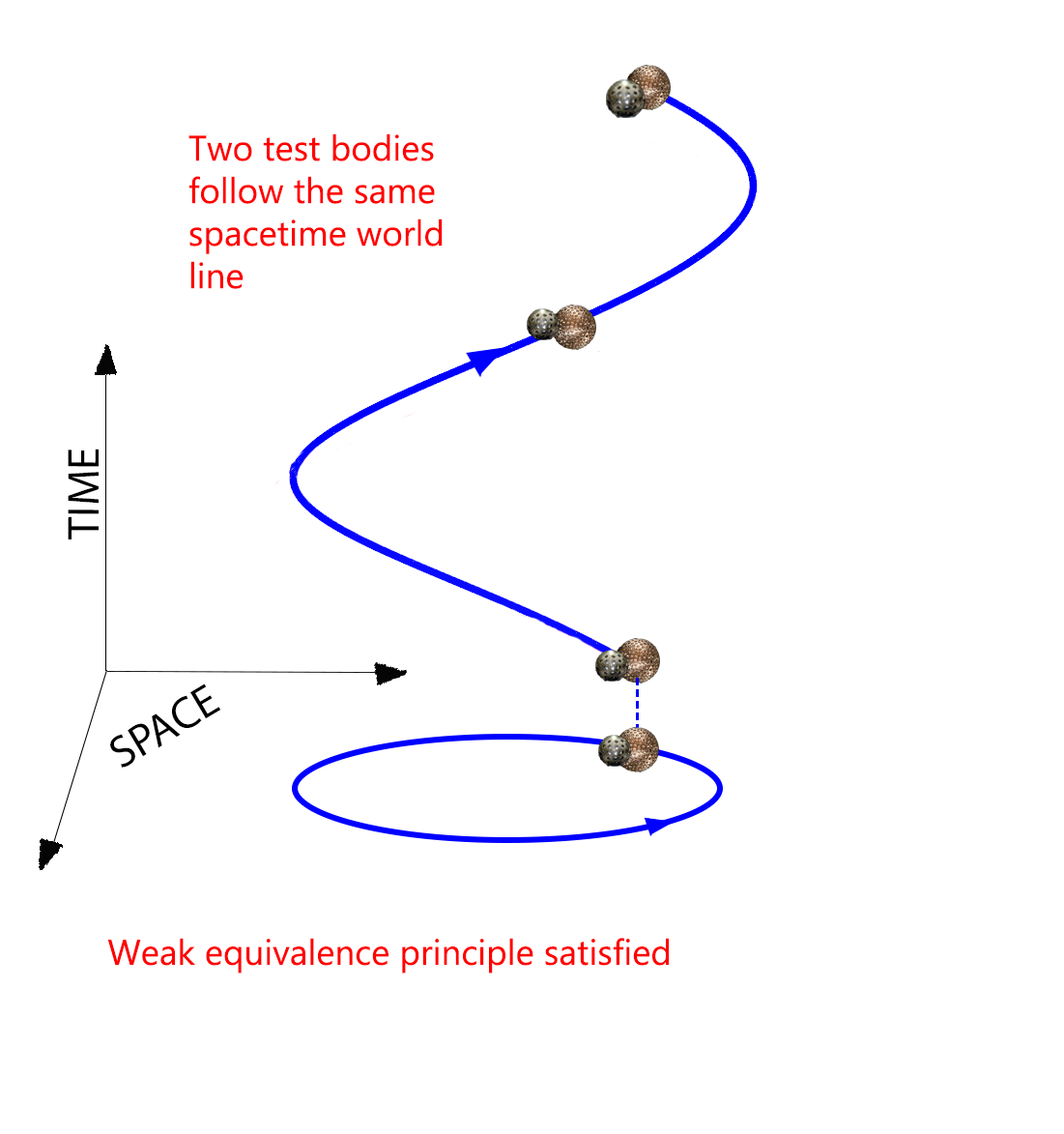}}\hfill
\subfloat[If there is a violation of the uniqueness of free fall, two bodies with the same initial conditions will not follow the same spacetime curves and their projections onto a spatial plane will, for example, be two different ellipses.\label{fig:1b}] {\includegraphics[width=0.45\textwidth]{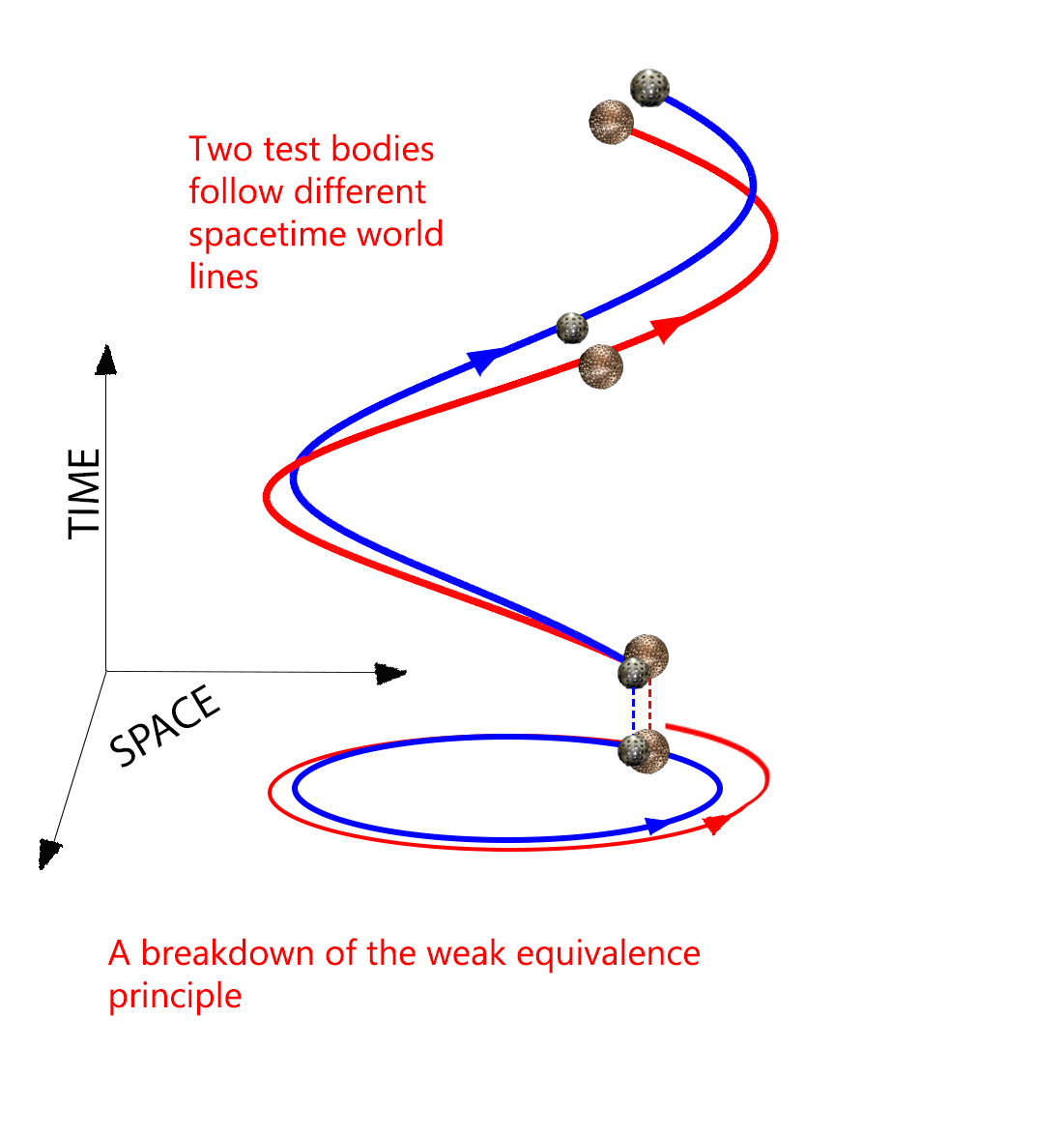}}\hfill

\caption{General Relativity and the Equivalence Principle.} \label{fig:1}
\end{figure}

\begin{figure}[htbp]
\centering
\subfloat[Residual radial accelerations of LAGEOS.\label{fig:2a}]{\includegraphics[width=0.45\textwidth]{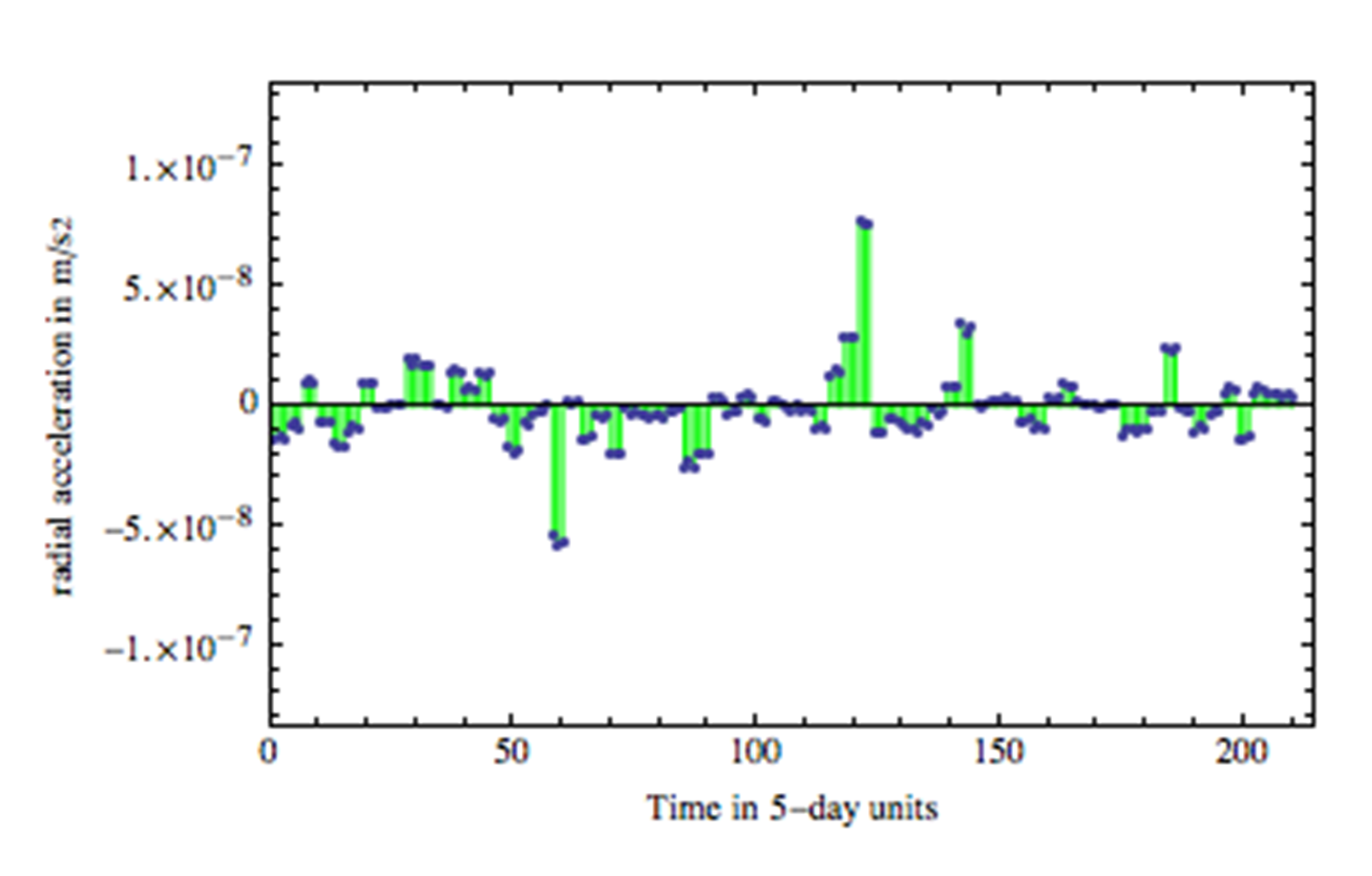}}\hfill
\subfloat[Residual radial accelerations of LAGEOS 2.\label{fig:2b}] {\includegraphics[width=0.45\textwidth]{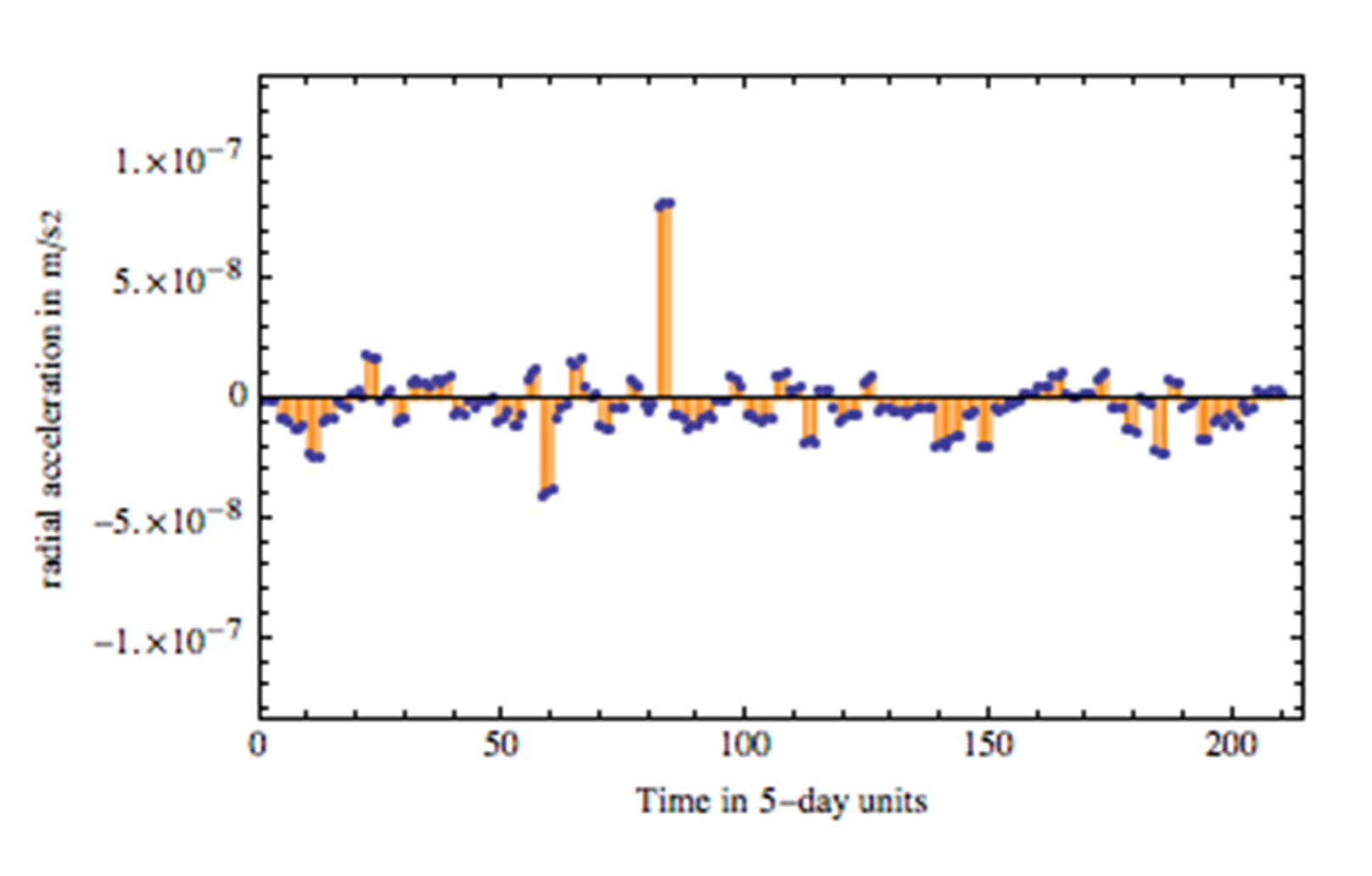}}\hfill
\subfloat[Residual radial accelerations of LARES.\label{fig:2b}] {\includegraphics[width=0.45\textwidth]{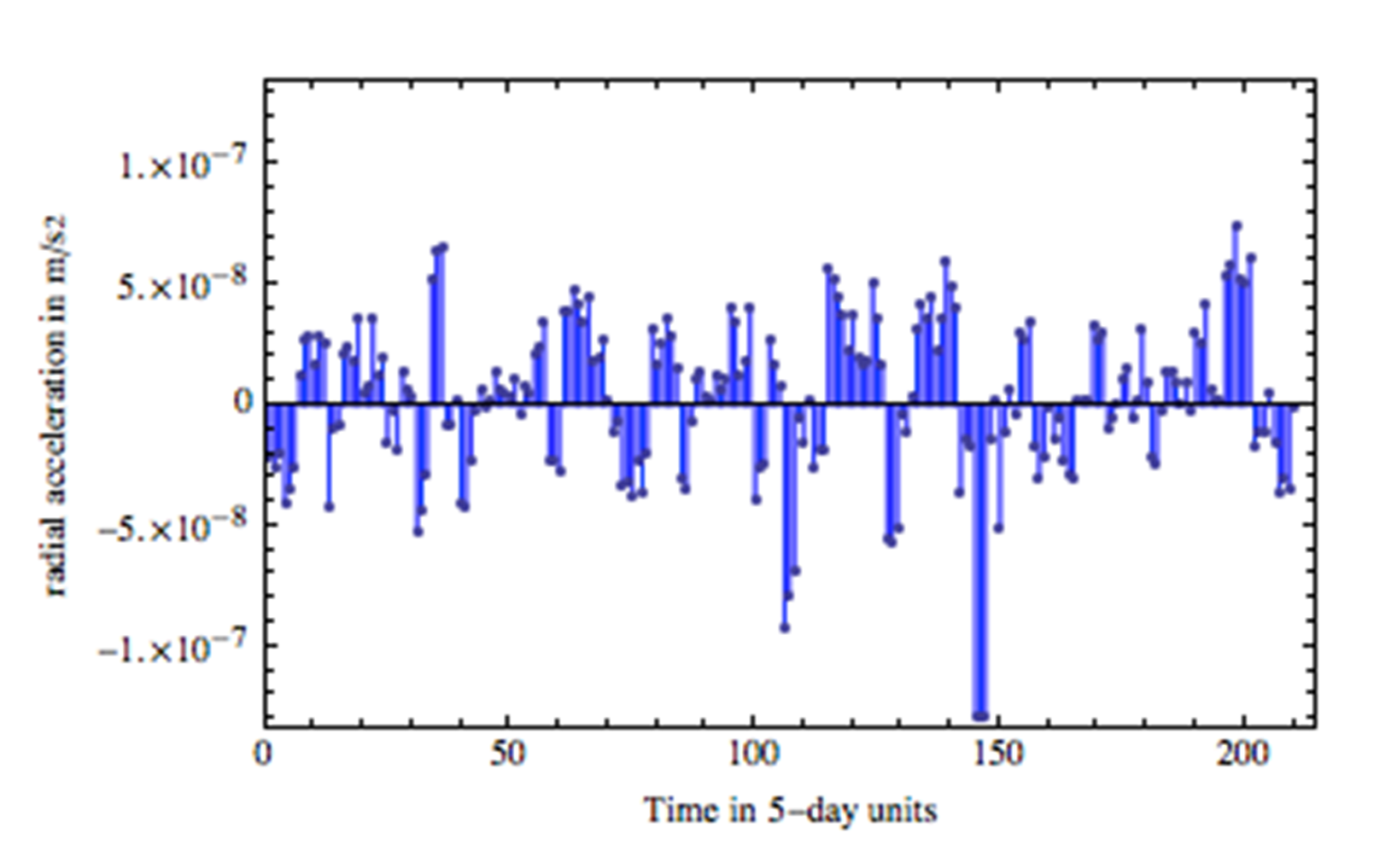}}\hfill
\caption{Residual radial accelerations.} \label{fig:2}
\end{figure}
%
%
%

\begin{figure}[H]
\includegraphics[width=4.5in]{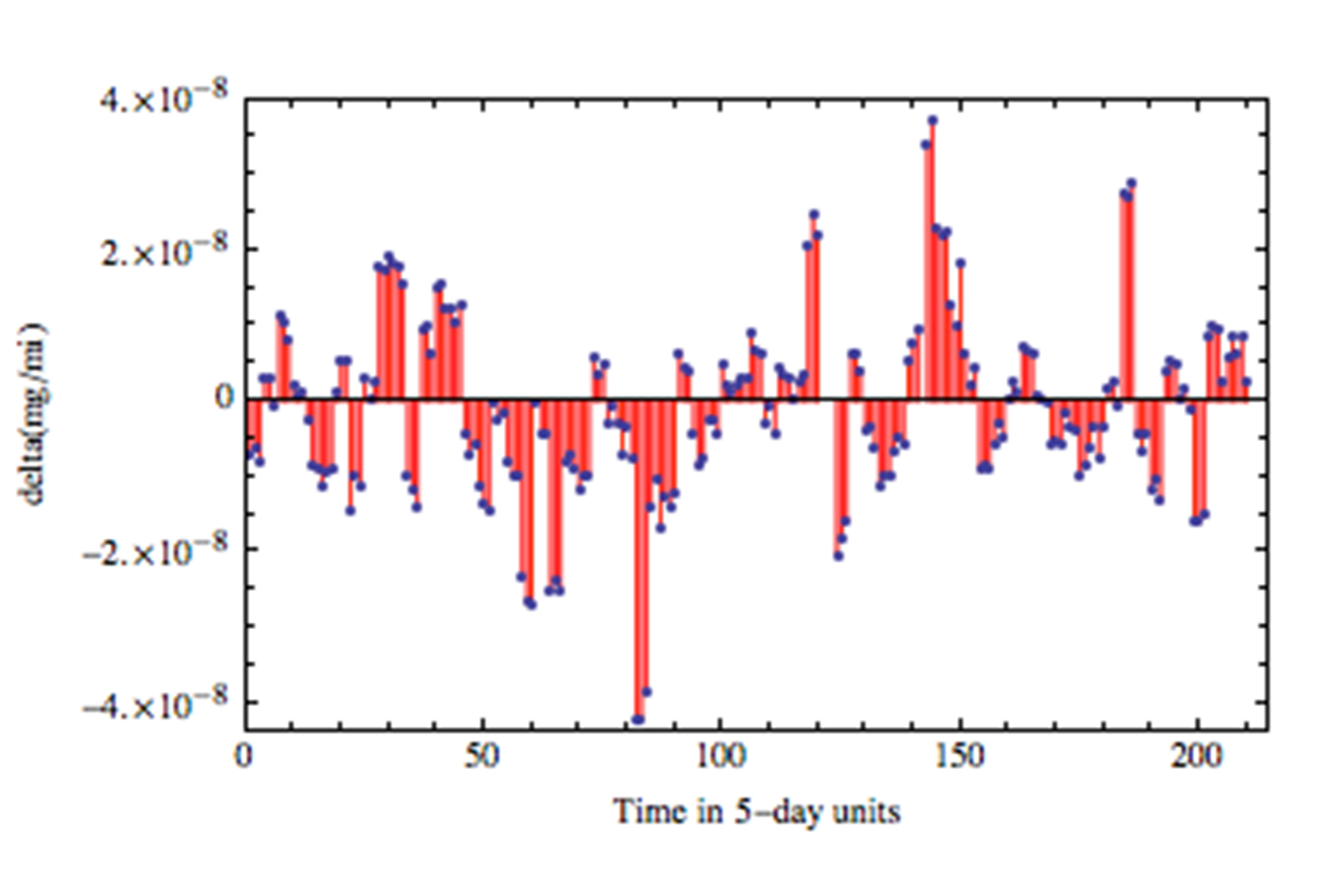} 
\caption{The residuals showing the deviation $\delta (m_{g}/m_{i} )$ for tungsten and aluminum/brass, Eq.(3), obtained by combining the residuals of the radial acceleration of the three satellites LARES, LAGEOS and LAGEOS 2. The variations over the mean are mainly due to the uncertainties in the variations of the Earth gravity field from spherical symmetry, i.e. to the uncertainties in the Earth spherical harmonics.}
\end{figure}

\section*{Author Contribution Statement}

I. C., E. C. P., G. S., J. R., and R. K. designed the data modeling and analysis, and carried out the precision orbit determination using the programs GEODYN, UTOPIA, and EPOSOC.

I. C.,  R. M., V. G., and A. P.  contributed to the experimental design, the theoretical interpretation and manuscript structure. 

All authors contributed to writing and editing the final manuscript.

\end{document}